\documentclass[fleqn,usenatbib,usedcolumn]{mnras}

\usepackage[british]{babel}             
\usepackage{amsmath}
\usepackage{newtxtext}                  
   \usepackage[slantedGreek]{newtxmath}    
   \usepackage[T1]{fontenc}                
   \usepackage{graphicx}                   
 \usepackage{multirow}

\renewcommand{\vec}[1]{\boldsymbol{#1}}

\newcommand{\xoff}{$x_{0}$}
\newcommand{\yoff}{$y_{0}$}
\newcommand{\thetas}{$\theta_{\rm s}$}
\newcommand{\Ytot}{$Y_{\rm tot}$}
\newcommand{\aGNFW}{$\alpha$}
\newcommand{\bGNFW}{$\beta$}
\newcommand{\cGNFW}{$\gamma$}
\newcommand{\xoffm}{x_{0}}
\newcommand{\yoffm}{y_{0}}
\newcommand{\thetasm}{\theta_{\rm s}}
\newcommand{\Ytotm}{Y_{\rm tot}}
\newcommand{\aGNFWm}{\alpha}
\newcommand{\bGNFWm}{\beta}
\newcommand{\cGNFWm}{\gamma}

\title[SZ profile fitting with joint AMI-\emph{Planck} analysis]{Sunyaev--Zel'dovich profile fitting with joint AMI-\emph{Planck} analysis}

\author[Y. C. Perrott et al.]{
Yvette C. Perrott$^{1,2}$\thanks{E-mail: yvette.perrott@vuw.ac.nz},
Kamran Javid$^{1}$,
Pedro Carvalho$^{3}$,
Patrick J. Elwood$^{1}$,
\newauthor
Michael P. Hobson$^{1}$,
Anthony N. Lasenby$^{1,4}$,
Malak Olamaie$^{1,5}$,
\newauthor
and Richard D. E. Saunders$^{1,4}$
\\
$^{1}$Astrophysics Group, Cavendish Laboratory, JJ Thomson Avenue, Cambridge CB3 0HE, UK\\
$^{2}$School of Chemical and Physical Sciences, Victoria University of Wellington, PO Box 600, Wellington 6140, New Zealand\\
$^{3}$Cambridge Machines Deep Learning and Bayesian Systems (CMDLABS) Ltd, 22 Wycombe End, Beaconsfield, Buckinghamshire, United Kingdom, HP9 1NB\\
$^{4}$Kavli Institute for Cosmology Cambridge, Madingley Road, Cambridge, CB3 0HA, UK\\
$^{5}$Imperial Centre for Inference and Cosmology (ICIC), Imperial College, Prince Consort Road, London SW7 2AZ, UK 
}

\date{Accepted XXX. Received YYY; in original form ZZZ}

\pubyear{2019}

\begin{document}
\label{firstpage}
\pagerange{\pageref{firstpage}--\pageref{lastpage}}
\maketitle

\begin{abstract}
We develop a Bayesian method of analysing Sunyaev-Zel'dovich measurements of galaxy clusters obtained from the Arcminute Microkelvin Imager (AMI) radio interferometer system and from the \textit{Planck} satellite, using a joint likelihood function for the data from both instruments. Our method is applicable to any combination of \emph{Planck} data with interferometric data from one or more arrays.  We apply the analysis to simulated clusters and find that when the cluster pressure profile is known a-priori, the joint dataset provides precise and accurate constraints on the cluster parameters, removing the need for external information to reduce the parameter degeneracy.  When the pressure profile deviates from that assumed for the fit, the constraints become biased.  Allowing the pressure profile shape parameters to vary in the analysis allows an unbiased recovery of the integrated cluster signal and produces constraints on some shape parameters, depending on the angular size of the cluster.  When applied to real data from \textit{Planck}-detected cluster PSZ2~G063.80+11.42, our method resolves the discrepancy between the AMI and \emph{Planck} $Y$-estimates and usefully constrains the gas pressure profile shape parameters at intermediate and large radii.

\end{abstract}

\begin{keywords}
methods: data analysis -- galaxies: clusters: general -- galaxies: clusters: individual: PSZ2~G063.80+11.42 -- galaxies: clusters: intracluster medium -- cosmology: observations.
\end{keywords}


\section{Introduction}

With the advent of large Sunyaev-Zel'dovich (SZ) effect surveys carried out by instruments such as \emph{Planck} \citep{2016A&A...594A..27P}, the Atacama Cosmology Telescope \citep{2018ApJS..235...20H}, and the South Pole Telescope \citep{2015ApJS..216...27B}, SZ observations have the potential to become a powerful tool for constraining, for example, cosmological properties via cluster number counts.  Numerical simulations show a tight, low-scatter correlation between the SZ observable, the Compton-$y$ parameter and cluster mass (e.g.\ \citealt{2004MNRAS.348.1401D}, \citealt{2006ApJ...650..538N}), but recent attempts to use SZ cluster number counts for cosmological analysis have produced results in tension with other, more mature methods such as the Cosmic Microwave Background (CMB) primary anisotropies \citep{2016A&A...594A..24P}.  One issue is the uncertain mass-observable calibration, although another issue is the modelling of the observable itself.

\citet{2015A&A...580A..95P} (hereafter P15) compared properties of 99 galaxy clusters observed in SZ with \emph{Planck} and the Arcminute Microkelvin Imager (AMI) radio interferometer system.  They showed that the discrepancies between the cluster parameters as constrained by AMI and \emph{Planck} could be explained by the cluster gas pressure profile deviating from the profile assumed for analysis.  The AMI observations were shown to be particularly sensitive to this effect when attempting to constrain the total integrated Compton-$y$ parameter due to missing angular scales.  It was noted in P15 that the combination of the two instruments would be powerful for investigating the gas pressure profiles of the clusters due to the complementary angular scales measured.  In this paper we explore this idea further by developing a joint Bayesian analysis pipeline which combines the data from the two instruments.  We note that this pipeline could also be used with other interferometric data, for example from the Atacama Large Millimeter/submillimeter Array (ALMA) which has recently been used for SZ analysis (e.g.\ \citealt{2016PASJ...68...88K}) and in the future for the Square Kilometre Array (SKA), which will be able to observe the SZ effect in its highest frequency band \citep{2015aska.confE.170G}.  The pipeline would also be easily extended to include data from other single-dish instruments such as NIKA(2) (e.g.\ \citealt{2014A&A...569A..66A}). This work joins a growing body of analysis combining SZ data from different instruments, sensitive to different angular scales.  Recent work such as \citet{2016ApJ...832...26S}, \citet{2017A&A...597A.110R}, and \citet{2018arXiv181201034D} all combine Planck Modified Internal Linear Combination Algorithm (MILCA; \citealt{2016A&A...594A..22P}) $y$-maps with SZ data from other instruments to jointly fit the gas pressure profile, while \citet{2018A&A...612A..39R} use a Planck-derived prior on the integrated Compton-$y$ parameter; however to our knowledge this is the first time that Planck frequency maps (rather than $y$-maps) have been jointly analysed with other SZ data.

The paper is arranged as follows.  In Section~\ref{sec:telescopes} we describe the telescopes and data used in our analysis and in Section~\ref{S:method} we describe our analysis method.  In Section~\ref{S:model} we describe our cluster model.  In Sections~\ref{S:sims} and ~\ref{S:sim_results} we verify our method using simulated clusters and in Section~\ref{S:real_data} we apply the method to a test case using real data.  We anticipate further work in Section~\ref{S:future_work} and conclude in Section~\ref{S:conclusions}.

\section{\textit{Planck} and AMI telescopes}
\label{sec:telescopes}


\subsection{\textit{Planck} satellite}

The combination of the \textit{Planck} satellite's low-frequency and high-frequency instruments (LFI and HFI) provides nine frequency channels in the range 37\,GHz -- 857\,GHz.  The HFI, used for cluster analysis, has angular resolutions of $10$, $7.1$, and $5.5$\,arcminutes at $100$, $143$, and $217$~GHz and $5.0$~arcminutes at each of $353$, $545$, and $857$\,GHz.  The \emph{Planck} frequency bands correspond to two decrements, the null, and three increments in the SZ spectrum, making it particularly effective at the blind identification of galaxy clusters despite its relatively low angular resolution.  See e.g.\ \citealt{2016A&A...594A..27P} for further details.


\subsection{AMI}
AMI \citep{2008MNRAS.391.1545Z} is a dual-array interferometer designed for SZ studies, which is situated near Cambridge, UK.  AMI consists of two arrays: the Small Array (SA), optimised for viewing arcminute-scale features, having an angular resolution of $\approx$\,3\,arcmin and sensitivity to structures up to $\approx$\,10\,arcmin in scale; and the Large Array (LA), with angular resolution of $\approx$\,30\,arcsec, which is insensitive to SA angular scales and is used to characterise and subtract confusing radio-sources.  Both arrays operate over the same frequency band with a central frequency of $\approx$\,15.5\,GHz and a bandwidth of $\approx$\,5\,GHz; in P15 this was divided into 6 channels but after a recent correlator upgrade the band is now divided into 4096 channels \citep{2018MNRAS.475.5677H}, and binned down to 8 channels for analysis after radio-frequency-interference excision and calibration.  The simulated data in this paper have the properties of the new correlator.


\section{Joint likelihood analysis}\label{S:method}

\subsection{Bayesian parameter estimation}

For a model, $\mathcal{M}$ and a data vector, $\vec{\mathcal{D}}$, we can obtain the probability distributions of model parameters (also known as input or sampling parameters) $\vec{\Theta}$ conditioned on $\mathcal{M}$ and $\vec{\mathcal{D}}$ using Bayes' theorem:
\begin{equation}\label{eqn:bayes}
Pr\left(\vec{\Theta}|\vec{\mathcal{D}},\mathcal{M}\right) = \frac{Pr\left(\vec{\mathcal{D}}|\vec{\Theta},\mathcal{M}\right)Pr\left(\vec{\Theta}|\mathcal{M}\right)}{Pr\left(\vec{\mathcal{D}}|\mathcal{M}\right)},
\end{equation}
where $Pr(\vec{\Theta}|\vec{\mathcal{D}},\mathcal{M}) \equiv \mathcal{P}(\vec{\Theta})$ is the posterior distribution of the model parameter set, $Pr(\vec{\mathcal{D}}|\vec{\Theta},\mathcal{M}) \equiv \mathcal{L}(\vec{\Theta})$ is the likelihood function for the data, $Pr(\vec{\Theta}|\mathcal{M}) \equiv \pi(\vec{\Theta})$ is the prior probability distribution for the model parameter set, and $Pr(\vec{\mathcal{D}}|\mathcal{M}) \equiv \mathcal{Z}(\vec{\mathcal{D}})$ is the Bayesian evidence of the data given a model $\mathcal{M}$.  In this paper we will be interested in the posterior distributions of the sampling parameters rather than the evidence, which would be used for model comparison.  We use the nested sampling algorithm \textsc{MultiNest} \citep{2009MNRAS.398.1601F} to calculate our posteriors.

\subsection{Model parameters}

The model parameters can be split into two subsets (which are assumed to be independent of one another): cluster parameters $\vec{\Theta}_{\rm cl}$ and radio-source or `nuisance' parameters $\vec{\Theta}_{\rm rs}$.
The cluster model parameters are relevant to both AMI and \textit{Planck} data, and are detailed with their associated priors $\pi(\vec{\Theta}_{\rm cl})$ in Section~\ref{S:model}.
$\vec{\Theta}_{\rm rs}$ are relevant only for AMI data, since they are used to model the radio-source contamination of the SZ cluster signal recorded by the SA, based on values measured with the LA. More information on the prior distributions used for $\vec{\Theta}_{\rm rs}$ can be found in Section~4.3 of P15.

\subsection{Joint likelihood function}
\label{s:ap_jlhood}

If one has an AMI dataset $\vec{d}_{\mathrm{AMI}}$ and a \textit{Planck} dataset $\vec{d}_{\mathrm{Pl}}$, then the \textit{joint} likelihood function for the data is given by
\begin{equation}
\label{e:ap_lhood}
\mathcal{L}(\vec{\Theta}) = \mathcal{L}\left(\vec{d}_{\mathrm{AMI}}, \vec{d}_{\mathrm{Pl}} |\vec{\Theta}, \mathcal{M}\right).
\end{equation}
In this analysis we treat $\vec{d}_{\mathrm{AMI}}$ and $\vec{d}_{\mathrm{Pl}}$ as being independent (see Section~\ref{s:ap_sims_cov} for justification), and since the \textit{Planck} predicted data only rely on the cluster parameters we can write
\begin{equation}
\label{e:ap_lhood2}
\mathcal{L}(\vec{\Theta}) = \mathcal{L}_{\mathrm{AMI}}\left(\vec{d}_{\mathrm{AMI}} | \vec{\Theta}, \mathcal{M} \right) \mathcal{L}_{\mathrm{Pl}} \left( \vec{d}_{\mathrm{Pl}} |\vec{\Theta}_{\rm cl}, \mathcal{M}\right).
\end{equation}

\subsubsection{AMI likelihood function}
\label{s:ap_AMIlhood}

The AMI likelihood calculation is detailed in \citet{2009MNRAS.398.2049F} (herafter F09).  Briefly, the AMI likelihood function $\mathcal{L}_{\mathrm{AMI}}\left(\vec{d}_{\mathrm{AMI}} | \vec{\Theta}, \mathcal{M} \right) \equiv \mathcal{L}_{\mathrm{AMI}}(\vec{\Theta})$ is given by
\begin{equation}\label{e:ap_AMIlhood} 
\mathcal{L}_{\mathrm{AMI}}(\vec{\Theta}) = \frac{1}{Z_{D}}e^{-\frac{1}{2}\chi_{\mathrm{AMI}}^{2}}.
\end{equation}
Here $\chi_{\mathrm{AMI}}^{2}$ is a measure of the goodness-of-fit between the real and modelled data and can be expressed as
\begin{equation}\label{eqn:chisq}
\begin{split}
\chi_{\mathrm{AMI}}^{2} = \sum\limits_{\nu,\nu'} \left [\vec{d}_{\mathrm{AMI},\nu} - \vec{d}_{\mathrm{AMI},\nu}^{\rm p}(\vec{\Theta}) \right ]^{\rm T} \mathbfss{C}_{\mathrm{AMI},\nu,\nu'}^{-1} \times \\
\left [ \vec{d}_{\mathrm{AMI},\nu'} - \vec{d}_{\mathrm{AMI},\nu'}^{\rm p}(\vec{\Theta}) \right ].
\end{split}
\end{equation} 
In this expression $\vec{d}_{\mathrm{AMI},\nu}$ are the data observed by AMI at frequency $\nu$, and $\vec{d}_{\mathrm{AMI},\nu}^{\rm p}(\vec{\Theta})$ are the predicted data generated by the model at the same frequency.
$\mathbfss{C}_{\mathrm{AMI},\nu,\nu'}$ is the theoretical covariance matrix for the AMI likelihood, which includes primordial CMB and source confusion noise as described in \citet{2002MNRAS.334..569H} and F09 (Section~5.3). Source confusion noise allows for the remaining radio-sources with flux densities below the flux limit $S_{\rm lim}$ that the LA can subtract down to. The instrumental noise is estimated from the scatter of the visibility measurements within an observation. 
Referring back to equation~\ref{e:ap_AMIlhood}, $Z_{D}$ is a normalisation constant given by $(2\pi)^{D / 2} |\mathbfss{C}_{\mathrm{AMI}}|^{1/2}$ where $D$ is the length of $\vec{d}_{\mathrm{AMI}}$ (i.e. the combined vector of data from all frequencies).

\subsubsection{\emph{Planck} likelihood function}
\label{s:ap_PwSlhood}

To calculate the \emph{Planck} likelihood, we use a version of the PowellSnakes (PwS; \citealt{2009MNRAS.393..681C} and \citealt{2012MNRAS.427.1384C}) Bayesian detection algorithm designed for detecting galaxy clusters in \emph{Planck} data, adapted to operate on a previously determined position rather than to conduct a blind search.  PwS treats the data observed by \textit{Planck} as a superposition of background sky emission (including foreground emission and primordial CMB) $\vec{b}_{\nu}$, instrumental noise $\vec{n}_{\nu}$, and signal from the (cluster) source $\vec{s}_{\nu}$. The model for the predicted data vector is thus
\begin{equation}
\label{e:ap_PwSd}
\vec{d}_{\mathrm{Pl},\nu}^{\rm p}(\vec{\Theta}_{\rm cl}) = \vec{s}_{\nu}(\vec{\Theta}_{\rm cl}) + \vec{b}_{\nu} + \vec{n}_{\nu}.
\end{equation}
PwS works with patches of sky sufficiently small such that it can be assumed the noise contributions are statistically homogeneous. In this limit it is more convenient to work in Fourier space, as the Fourier modes are uncorrelated.  It also assumes that the noise contributions are Gaussian, which is very accurate in the case of instrumental noise and the primordial CMB but may be more questionable for Galactic emission -- the deviations from Gaussianity of $\vec{b}_{\nu}$ are discussed in Section~4.3 of \citealt{2012MNRAS.427.1384C}. 
Since PwS is a detection algorithm, it calculates the ratio of the likelihood of detecting a cluster parameterised by $\vec{\Theta}_{\rm cl}$ and the likelihood of the data with no cluster signal ($\vec{s}_{\nu}(\vec{\Theta}_{\rm cl, 0}$) = 0). Thus the log-likelihood ratio 
of the Fourier transformed quantities is 
\begin{equation}
\label{e:ap_PwSlhoodr}
\begin{split}
\ln \left[ \frac{\mathcal{L}_{\mathrm{Pl}}\left(\vec{\Theta}_{\rm cl}\right)}{\mathcal{L}_{\mathrm{Pl}}\left(\vec{\Theta}_{\rm cl,0}\right)} \right] = & \sum\limits_{\nu,\nu'} \tilde{\vec{d}}_{\mathrm{Pl},\nu}^{\rm p}(\vec{\Theta}_{\rm cl})^{\rm T} \mathbfss{C}_{\mathrm{Pl},\nu,\nu'}^{-1} \tilde{\vec{d}}_{\mathrm{Pl},\nu}(\vec{\Theta}_{\rm cl}) \\
 & - \frac{1}{2} \tilde{\vec{d}}_{\mathrm{Pl},\nu}^{\rm p}(\vec{\Theta}_{\rm cl})^{\rm T} \mathbfss{C}_{\mathrm{Pl},\nu,\nu'}^{-1} \tilde{\vec{d}}_{\mathrm{Pl},\nu}^{\rm p}(\vec{\Theta}_{\rm cl}),
\end{split}
\end{equation}
where tildes denote the Fourier transform of a quantity, and $\mathbfss{C}_{\mathrm{Pl},\nu,\nu'}$ is the covariance matrix of the data in Fourier space.

\subsubsection{Joint likelihood analysis hyperparameters}

Ideally when making a joint inference from two different datasets, likelihood hyperparameters would be used (see \citealt{2000MNRAS.315L..45L}, \citealt{2002MNRAS.335..377H}, and \citealt{2014A&C.....5...45M}) so that the relative weighting of the two likelihoods is treated in a Bayesian way. This allows meaningful results to be extracted when the datasets are not in good agreement, which could be due to, for example, systematic bias or an incorrect model.  However the log-ratio given by equation~\ref{e:ap_PwSlhoodr} is not a probability density due to the fact that it is missing a normalisation factor proportional to $\tilde{\vec{d}}_{\mathrm{Pl},\nu}(\vec{\Theta}_{\rm cl})^{\rm T} \mathbfss{C}_{\mathrm{Pl},\nu,\nu'}^{-1} \tilde{\vec{d}}_{\mathrm{Pl},\nu}(\vec{\Theta}_{\rm cl})$. These hyperparameters affect the normalisation factor of the likelihood (it becomes a function of them), thus the PwS algorithm is incompatible with their use as it does not include the normalisation term in the likelihood calculations (a more detailed explanation is given in Chapter~8 of \citealt{kj_thesis}). We therefore do not include likelihood hyperparameters in our analysis.  This should not be problematic when applying our method unless the data have been incorrectly analysed or the cluster model used is not flexible enough to describe both of the datasets.


\section{Cluster model}\label{S:model}

Since AMI and \textit{Planck} observe the SZ effect caused by the electron gas in clusters (see e.g. \citealt{1999PhR...310...97B}), both telescopes measure signals proportional to the Comptonisation parameter $y$,
\begin{equation}\label{e:ap_comptparam}
y = \frac{\sigma_{\rm T}k_{\rm B}}{m_{\rm e}c^{2}}\int T_{\rm e}(r) n_{\rm e}(r)\,\mathrm{d}l,
\end{equation}
where $k_{\rm B}$ is the Boltzmann constant, $m_{\rm e}$ is the rest mass of an electron, $c$ is the speed of light, and $\sigma_{\rm T}$ is the Thomson scattering cross-section. $T_{\rm e}(r)$ and $n_{\rm e}(r)$ are respectively the electron temperature and number density in the intra-cluster medium, as a function of radius from the centre of the cluster ($r$), and the integral is over the line of sight.
If an ideal gas equation of state is assumed for the electron gas then in terms of the electron pressure $P_{\mathrm{e}}(r)$, the Comptonisation parameter is given by
\begin{equation}\label{e:ap_comptparamideal}
y = \frac{\sigma_{\rm T}}{m_{\rm e}c^{2}}\int P_{\rm e}(r)\,\mathrm{d}l.
\end{equation}
The cluster model considered here is used to calculate a `map' of $y$ by evaluating equation~\ref{e:ap_comptparamideal} at different points on the plane of the sky.  It assumes a spherically symmetric, generalised-Navarro-Frenk-White (GNFW, \citealt{2007ApJ...668....1N}) profile to model the electron pressure 
\begin{equation}\label{e:ap_gnfw}
P_{\rm e}(r) = \frac{P_{\rm ei}}{\left(\frac{r}{r_{\rm p}}\right)^{\cGNFWm}\left(1+\left(\frac{r}{r_{\rm p}}\right)^{\aGNFWm}\right)^{(\bGNFWm-\cGNFWm)/\aGNFWm}}.
\end{equation}
$P_{\rm ei}$ is an overall pressure normalisation factor and $r_{\rm p}$ is a characteristic radius.

The parameters \aGNFW, \bGNFW\ and \cGNFW\ describe the slope of the pressure profile at $ r \approx r_{\rm p}$, $r \gg r_{\rm p}$ and $r \ll r_{\rm p}$ respectively. Our model parameterises a cluster in terms of observational (rather than physical) quantities: \thetas, the characteristic angular scale corresponding to $r_{\rm p}$ ($\thetasm = r_{\rm p} / D_{A}$ where $D_{A}$ is angular diameter distance) and \Ytot, the total integrated Comptonisation parameter of the cluster, given by

\begin{equation}\label{eq:Ytot}
\Ytotm = \frac{4\pi\sigma_{\mathrm{T}}}{m_{\rm e}c^{2}} P_{\rm ei} D_{A} \thetasm^{3} \frac{\Gamma \left ( \frac{3-\gamma}{\alpha} \right ) \Gamma \left ( \frac{\beta-3}{\alpha} \right )}{\alpha \Gamma \left ( \frac{\beta-\gamma}{\alpha} \right )}
\end{equation}
where $\Gamma$ is the gamma function.

Our model therefore has input parameters $\vec{\Theta}_{\rm cl} = \left(\xoffm, \yoffm, \Ytotm, \thetasm, \aGNFWm, \bGNFWm, \cGNFWm \right)$, where \xoff\ and \yoff\ are the cluster centre offsets from the designated central sky coordinate.

A set of `universal' pressure profile (UPP) GNFW shape parameter values were derived in \citet{2010A&A...517A..92A} as the best fit to a sample of clusters from REXCESS (observed with XMM-Newton, \citealt{2007A&A...469..363B}).  These are $(\cGNFWm, \aGNFWm, \bGNFWm, c_{500})$ = (0.3081, 1.0510, 5.4905, 1.177) and are often used as a fixed standard cluster profile.  In this analysis we will not restrict our model profiles to the UPP case.

The GNFW profile extends to infinity; in practice some cut-off radius for the $y$-map must therefore be defined when implementing this model.  A frequent choice, used for example in the analysis of \emph{Planck} data, is to cut off at $\theta = 5\theta_{500} = 5\thetasm c_{500}$.  For the case of the UPP, this implies that a spherical integral to $5\theta_{500}$ gives $Y_{\mathrm{sph},5\theta_{500}} = 0.96\,\Ytotm$, or a cylindrical integral gives $Y_{\mathrm{cyl},5\theta_{500}} = 0.97\,\Ytotm$ (with the line-of-sight integral extending to infinity within a radius of $5\theta_{500}$ on the sky).  In the case of arbitrary values of $(\cGNFWm, \aGNFWm, \bGNFWm)$, this fraction can change significantly; plus in the general case $c_{500}$ is not necessarily known so a different cut-off radius must be defined which we denote $\theta_{\mathrm{lim}}$.  We choose to define $\theta_{\mathrm{lim}}$ for an arbitrary profile by the radius at which $Y_{\mathrm{sph},\theta_{\mathrm{lim}}} = 0.95\,\Ytotm$ (found via a numerical root-finder) by analogy with the UPP.  For some profiles which fall off more slowly with radius this becomes prohibitively large and we therefore impose a maximum radius of $10\,\thetasm$.  We have verified that in these cases the Comptonization parameter integrated over the line-of-sight at $10\,\thetasm$ is $<0.1$\% of the value at the centre, i.e.\ we are not cutting off substantial cluster signal in the outskirts. 

We then translate the $y$-map to signal on the sky at each frequency using the non-relativistic approximation \citep{1969Ap&SS...4..301Z}.  Recently it has been shown that relativistic corrections may be significant for high-temperature clusters at \emph{Planck} frequencies (e.g.\ \citealt{2019MNRAS.483.3459R}); we note that relativistic effects \emph{reduce} the \emph{Planck} SZ signal, so if corrections were applied the effect would be to \emph{increase} the discrepancy between AMI and \emph{Planck} parameter constraints shown in P15.

\subsection{Priors}

In P15 we used a position prior based on the \emph{Planck} position and error, and a joint ellisoidal prior on \Ytot\ and \thetas\ based on the population of clusters detected by \emph{Planck}.  Here we assign wide, non-informative, independent priors to \xoff, \yoff, \Ytot\ and \thetas\ (see Table~\ref{t:ap_priors}), to explore how much the combination of the two datasets can constrain the parameters. 

In the standard \emph{Planck} analysis and for the AMI data analysed in P15, the GNFW shape parameter values were fixed to the UPP values.  In this analysis we will both simulate and analyse clusters with non-UPP profiles and explore the constraints on \aGNFW\ and \bGNFW\ produced by our joint dataset.  The priors used for \aGNFW\ and \bGNFW\ vary throughout our analysis (Table~\ref{t:ap_priors}); they are either fixed at some specific value (delta function priors) or allowed to vary uniformly (uniform priors on a bounded domain).

\begin{table}
\begin{center}
\caption{Cluster parameter prior distributions. $\mathcal{N}$ denotes a Gaussian distribution parameterised by its mean and standard deviation, $\mathcal{U}$ denotes a uniform distribution, and $\delta$ is a Dirac delta function. In the cases where the latter is used, the values used for the function's argument will be stated when the analyses are carried out.}
\label{t:ap_priors}
\begin{tabular}{{l}{c}}
\hline
Parameter & Prior distribution(s) \\ 
\hline
\xoff & $\mathcal{N}(0'', 300'')$ \\
\yoff & $\mathcal{N}(0'', 300'')$ \\
\Ytot & $\mathcal{U}[0.00~\mathrm{arcmin}^{2}, 0.02~\mathrm{arcmin}^{2}]$ \\
\thetas & $\mathcal{U} [1.3', 15']$ \\
\aGNFW & $\delta(\aGNFWm_{\mathrm{model}})$ or $\mathcal{U}[0.1, 3.5]$ \\
\bGNFW & $\delta(\bGNFWm_{\mathrm{model}})$ or $\mathcal{U}[3.5, 7.5]$ \\
\cGNFW & $\delta(\cGNFWm_{\mathrm{model}})$ \\
\hline
\end{tabular}
\end{center}
\end{table}


\section{Cluster simulations}\label{S:sims}

For all simulations the $y$ map of a single cluster is generated with ten distinct noise realisations as follows.  Firstly, ten CMB realisations are created by sampling primordial CMB noise from an empirical power-law distribution \citep{2013ApJS..208...19H} and distributing at random positions on the sky.  For each instrument, we add further realisations of the relevant sources of noise, as follows.

\subsection{\emph{Planck} cluster simulations}
We construct a \emph{Planck} all-sky foreground and thermal noise map by adding one noise realisation to the CO, FIRB, free-free, spinning dust, synchrotron and thermal dust emission maps simulated using the \emph{Planck} sky model, all taken from the \emph{Planck} Legacy Archive\footnote{\url{https://pla.esac.esa.int/\#home}} using the FFP8 release.  We note that we do not include point sources since as of this data release they were not separated into strong and weak point source maps, and for cluster analysis on real data the strong point sources would be masked.  We also choose not to add the thermal and kinetic SZ emission maps as we wish to see how much information can be extracted from the data in an ideal situation of a single, isolated cluster.

We then randomly select ten patch centres on the sky, with the constraint that the patch centres are above $\delta = 0^{\circ}$ (to satisfy AMI's observing limits) and that all of the 20$^{\circ}$ square patch is outside of the \emph{Planck} 20\% Galactic plane mask (in which 20\% of the sky is masked).  We cut patches from the all-sky foreground $+$ noise map at these coordinates and add the resulting patch maps to the CMB and cluster maps to produce the final \emph{Planck} simulations.  Each patch map therefore contains different thermal noise, foreground emission and CMB but the same cluster.

\subsection{AMI cluster simulations}
The AMI simulations are constructed by adding the cluster and CMB maps to confusion noise realisations, created using the 10C source counts given in \citet{2011MNRAS.415.2708A} up to a maximum flux density of 360\,$\upmu$Jy (i.e.\ assuming sources above $4\times$ a typical AMI-LA noise limit of 90\,$\upmu$Jy/beam have been detected and removed). A ten-hour mock AMI-SA observation is performed at the ten patch centres used for the \emph{Planck} simulations, using the in-house package \textsc{Profile} (see e.g. \citealt{2002MNRAS.333..318G}).  Instrumental noise is also added to the mock observations, giving a total noise level of $\approx$\,120\,$\upmu$Jy/beam on the map.

Similarly to the \emph{Planck} simulations, each AMI simulation contains different thermal noise, confusion noise and CMB and the same cluster.  Each AMI simulation corresponds directly to a \emph{Planck} simulation which has the same observing centre and CMB realisation.

\subsection{Testing the independence of the AMI and \textit{Planck} datasets}
\label{s:ap_sims_cov}

\begin{figure}
  \begin{center}
     \includegraphics[width=\linewidth, keepaspectratio]{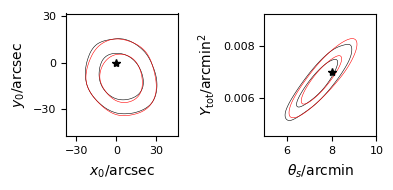}
\caption{Two-dimensional marginalised \xoff-\yoff\ and \Ytot-\thetas\ posterior distributions for a high SNR (see Section~\ref{S:sim_results}) cluster simulation. The red contours correspond to the posterior distribution associated with the AMI and \textit{Planck} datasets which had different CMB realisations to each other, while the black ones correspond to datasets generated with the same realisation. The star symbols indicate the values input when generating the simulations.}
\label{f:ap_cov_test}
  \end{center}
\end{figure}

In Section~\ref{s:ap_jlhood} we made the assumption that $\vec{d}_{\mathrm{AMI},\nu}^{\rm p}$ and $\vec{d}_{\mathrm{Pl},\nu}^{\rm p}$ are not correlated with each other, so that the likelihoods for the two datasets can be separated. The instrumental noises associated with each telescope can safely be assumed to be independent. Due to the telescopes operating at different angular scales and frequencies, the confusion noise present in AMI data and foreground emission present in \textit{Planck} data are assumed to be independent of one another. A similar argument can be applied for primordial CMB noise, nevertheless we carried out a simple test to see if this is the case. For a given set of cluster parameters, we ran the joint analysis on \textit{Planck} and AMI datasets which had different CMB realisations to one another. We found that the resultant parameter constraints were not significantly different to the results obtained using AMI and \textit{Planck} data which had the same CMB realisations as one another (Figure~\ref{f:ap_cov_test}). We thus concluded that the covariance between the datasets introduced by the common CMB background was negligible.

\section{Cluster simulation results}\label{S:sim_results}

In the following we apply the joint analysis to clusters simulated as described in Section~\ref{S:sims}, and compare results with analyses which use (the same) AMI or \textit{Planck} data alone.

We consider simulations generated using three different sets of profile shape parameters: the UPP, the parameters fitted to a stacked \emph{Planck} dataset in \citet{2013A&A...550A.131P} (`\textit{Planck} Intermediate Profile', PIP), and the parameters fitted to the cluster RXC\,J2319.6-7313 in \citet{2010A&A...517A..92A} which are the most different in the sample to the UPP (`REXCESS Extreme Profile', REP).  These values are listed in Table~\ref{t:ap_obs_sim}.  We consider a `low' and a `high' signal-to-noise ratio (SNR) cluster, which correspond to input values of $(\Ytotm, \thetasm) = (0.001\,\mathrm{arcmin}^{2}, 2\,\mathrm{arcmin})$ and $(\Ytotm, \thetasm) = (0.007\,\mathrm{arcmin}^{2}, 8\,\mathrm{arcmin})$ respectively. We note that `low' and `high' SNR are in relation to the \emph{Planck} simulations rather than the AMI simulations, where they are both well detected.  We then analyse the ten different noise realisations for each cluster using the priors given in Table~\ref{t:ap_priors} and plot the resulting posterior distributions using \textsc{GetDist}\footnote{\url{http://getdist.readthedocs.io/en/latest/}.}.

\begin{table}
\begin{center}
\caption{Cluster simulation inputs using an observational model and three different pressure profile shapes.  \xoff\ and \yoff\ are always 0, i.e.\ the cluster is at the simulated map centre.}
\label{t:ap_obs_sim}
\footnotesize
\begin{tabular}{lcccccc}
\hline
 & & \Ytot & \thetas & \cGNFW & \aGNFW & \bGNFW \\
 & & / arcmin$^2$ & / arcmin & & & \\
\hline
\multirow{2}{*}{UPP} & low SNR & 0.001 & 2 & \multirow{2}{*}{0.3081} & \multirow{2}{*}{1.0510} & \multirow{2}{*}{5.4905} \\
 & high SNR & 0.007 & 8 & & & \\
\multirow{2}{*}{PIP} & low SNR & 0.001 & 2 & \multirow{2}{*}{0.31} & \multirow{2}{*}{1.33} & \multirow{2}{*}{4.13} \\
 & high SNR & 0.007 & 8 & & & \\
\multirow{2}{*}{REP} & low SNR & 0.001 & 2 & \multirow{2}{*}{0.065} & \multirow{2}{*}{0.33} & \multirow{2}{*}{5.49} \\
 & high SNR & 0.007 & 8 & & & \\
\hline
\end{tabular}
\end{center}
\end{table}

\subsection{Analysis with fixed profile parameters}

\begin{figure*}
  \begin{center}
    \begin{tabular}{@{}c@{}}
     \includegraphics[width=\textwidth, height = 7.5cm, keepaspectratio]{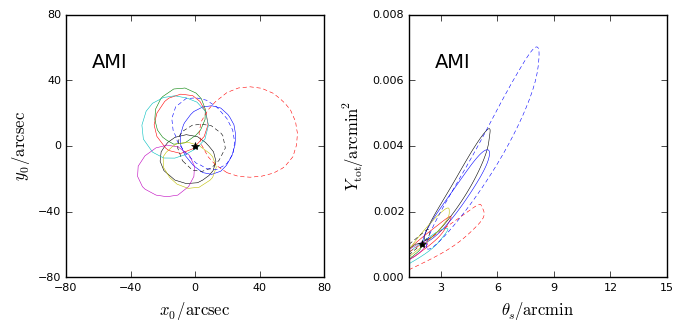} \\
     \includegraphics[width=\textwidth, height = 7.5cm, keepaspectratio]{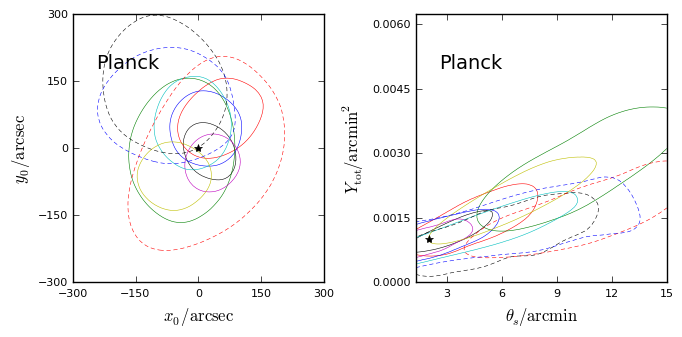} \\
     \includegraphics[width=\textwidth, height = 7.5cm, keepaspectratio]{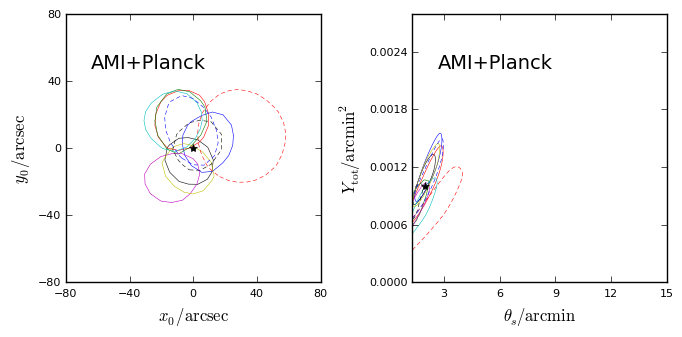} \\
    \end{tabular}
\caption{Two-dimensional marginalised \xoff-\yoff\ and \Ytot-\thetas\ posterior distributions for the 10 UPP low SNR cluster simulations obtained from: AMI data (top row), \textit{Planck} data (middle row), and AMI and \textit{Planck} data combined (bottom row). The contours in each plot represent the $68\%$ confidence intervals of the separate posterior distributions obtained from each of the 10 realisations. The star symbols indicate the values input when generating the simulations.  Note the different axis scales.}
\label{f:ap_obs_lowsnr_sims}
  \end{center}
\end{figure*}

\begin{figure*}
  \begin{center}
    \begin{tabular}{@{}c@{}}
     \includegraphics[width=\textwidth, height = 7.5cm, keepaspectratio]{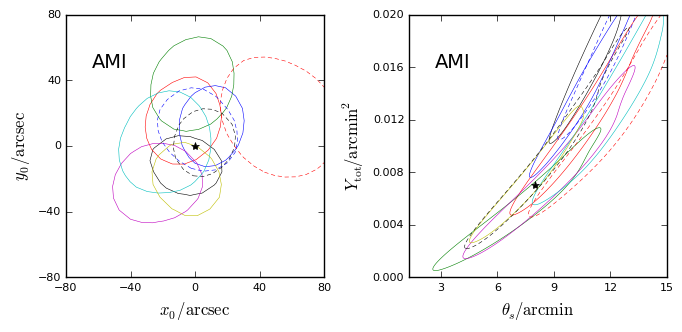} \\
     \includegraphics[width=\textwidth, height = 7.5cm, keepaspectratio]{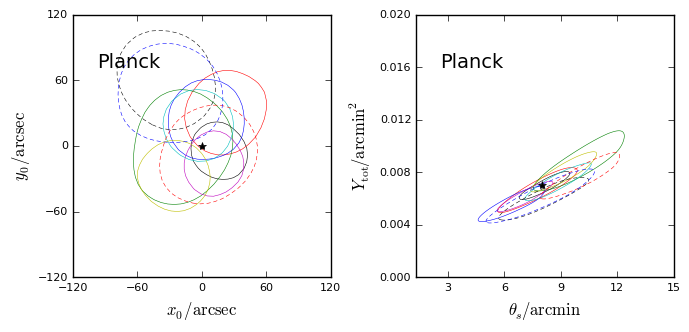} \\
     \includegraphics[width=\textwidth, height = 7.5cm, keepaspectratio]{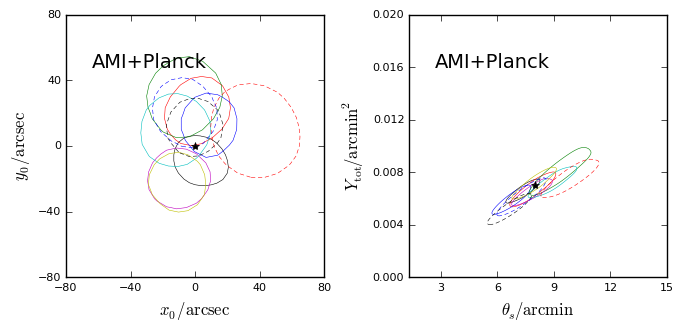} \\
    \end{tabular}
\caption{Two-dimensional marginalised \xoff-\yoff\ and \Ytot-\thetas\ posterior distributions for the 10 UPP high SNR cluster realisations. The Figure layout is as described in Figure~\ref{f:ap_obs_lowsnr_sims}.  Note the different axis scales in the \xoff-\yoff\ plots.}
\label{f:ap_obs_highsnr_sims}
  \end{center}
\end{figure*}

We firstly analyse the high- and low-SNR UPP simulations using $(\cGNFWm, \aGNFWm, \bGNFWm)$ fixed to the correct, input values.  Figs.~\ref{f:ap_obs_lowsnr_sims} and \ref{f:ap_obs_highsnr_sims} show the results of the simulation sets using AMI data only, \emph{Planck} data only, and the two datasets combined.  Each two-dimensional marginalised posterior distribution plot shows the 68\% confidence contours for the 10 different noise realisations, with the input values marked with black stars.

In the case of the positional parameters, it is clear that the higher angular resolution of the AMI data means that it drives the posterior inferences, although the addition of the \emph{Planck} data does improve the constraints slightly in the high-SNR case.  This result is consistent for all following analyses, and we do not consider the \xoff\ and \yoff\ constraints any further.

Both AMI and \emph{Planck} constraints are degenerate in the \thetas/\Ytot\ plane and previous results based on AMI and \emph{Planck} data have relied on ancillary data to reduce this degeneracy.  For AMI, a correlated prior in \thetas\ and \Ytot\ based on a simulated population of clusters injected into and recovered from \emph{Planck} data was used (see P15).  For \emph{Planck}, a prior on \thetas\ has been used to `slice' the $\theta_{500}$/$Y(r_{500})$ posterior constraint based on either an X-ray measurement or a mass-observable scaling relationship (see, e.g.\ \citealt{2016A&A...594A..27P}), both relying on the assumption of the `universal' $c_{500}$.  The combination of AMI and \emph{Planck} data removes the need for these ancillary priors and produces a much tighter constraint on both \thetas\ and \Ytot\ since the degeneracy directions are different.  This is most striking in the case of the low SNR cluster, but is also evident in the case of the high SNR cluster.

\begin{figure*}
  \begin{center}
    \begin{tabular}{@{}ccc@{}}
     \includegraphics[width=0.3\textwidth, keepaspectratio]{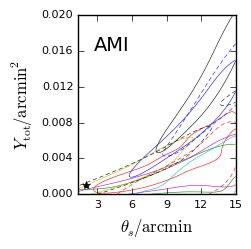} & \includegraphics[width=0.3\textwidth, keepaspectratio]{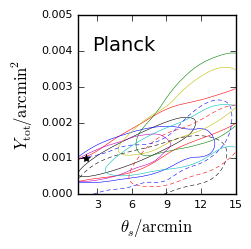} & \includegraphics[width=0.3\textwidth, keepaspectratio]{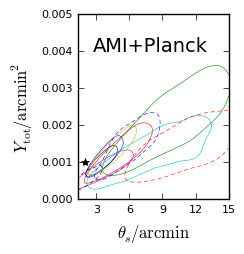} \\
     \includegraphics[width=0.3\textwidth, keepaspectratio]{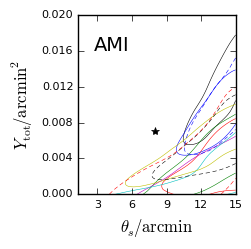} & \includegraphics[width=0.3\textwidth, keepaspectratio]{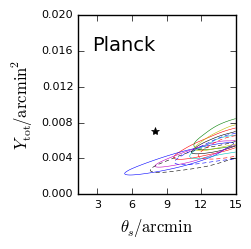} & \includegraphics[width=0.3\textwidth, keepaspectratio]{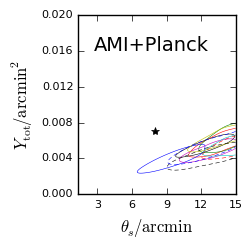} \\
    \end{tabular}
\caption{Two-dimensional marginalised \Ytot-\thetas\ posterior distributions for the 10 PIP low SNR cluster simulations (top row) and high SNR cluster simulations (bottom row).  The posteriors plotted are, from left to right: AMI-only, \emph{Planck}-only, and joint constraints. The contours and markers are as described in Figure~\ref{f:ap_obs_lowsnr_sims}.  Note the different axis scales in the low-SNR plots.}
\label{f:pip_fixed}
  \end{center}
\end{figure*}

\begin{figure*}
  \begin{center}
    \begin{tabular}{@{}ccc@{}}
     \includegraphics[width=0.3\textwidth, keepaspectratio]{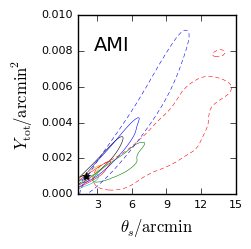} & \includegraphics[width=0.3\textwidth, keepaspectratio]{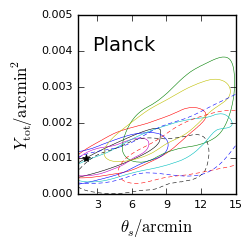} & \includegraphics[width=0.3\textwidth, keepaspectratio]{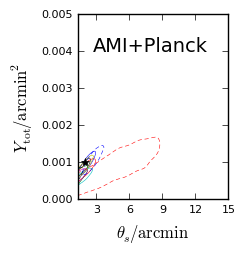} \\
     \includegraphics[width=0.3\textwidth, keepaspectratio]{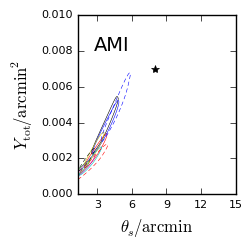} & \includegraphics[width=0.3\textwidth, keepaspectratio]{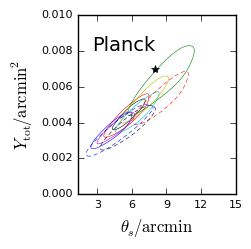} & \includegraphics[width=0.3\textwidth, keepaspectratio]{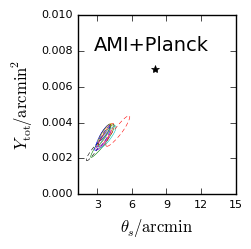} \\
    \end{tabular}
\caption{Two-dimensional marginalised \Ytot-\thetas\ posterior distributions for the 10 REP low SNR cluster simulations (top row) and high SNR cluster simulations (bottom row).  The posteriors plotted are, from left to right: AMI-only, \emph{Planck}-only, and joint constraints. The contours and markers are as described in Figure~\ref{f:ap_obs_lowsnr_sims}.  Note the different y-axis scales.}
\label{f:rep_fixed}
  \end{center}
\end{figure*}

We next analyse the PIP and REP simulations keeping $(\cGNFWm,\aGNFWm,\bGNFWm)$ fixed to the UPP values in the analysis, i.e.\ we now have a mismatch between the cluster profiles used to produce and analyse the simulations.  The two-dimensional posterior constraints on \Ytot\ and \thetas\ in this case are shown in Figs.~\ref{f:pip_fixed} and \ref{f:rep_fixed}.  In the case of the PIP simulations, all of the constraints are offset from the true position; although the direction of the offset is mostly in \thetas\ so that the one-dimensional marginal constraint on \Ytot\ is not too badly offset it is clear that any method to reduce the degeneracy by slicing the posterior will be problematic.  In the case of the REP simulations, the low-SNR cluster constraints are not too badly affected by the profile mismatch; this is because the change is to the inner part of the profile which is not well-resolved by either instrument.  The high-SNR constraints however are significantly biased both in the two-dimensional plane and in the one-dimensional \Ytot\ plane.

\subsection{Variable shape parameter analysis}
\label{s:ap_obs_highsnrab_sims}
We next consider the same set of simulations described in Section~\ref{S:sim_results}, but now allowing the GNFW shape parameters \aGNFW\ and \bGNFW\ to vary in the \textit{analysis}. We assign the uniform priors stated in Table~\ref{t:ap_priors} to \aGNFW\ and \bGNFW.

\begin{figure*}
  \begin{center}
    \begin{tabular}{@{}c@{}}
     \includegraphics[width=\textwidth, height = 7.5cm, keepaspectratio]{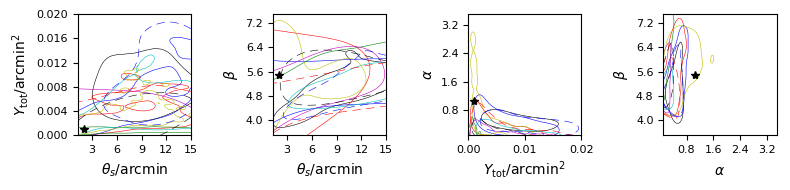} \\
     \includegraphics[width=\textwidth, height = 7.5cm, keepaspectratio]{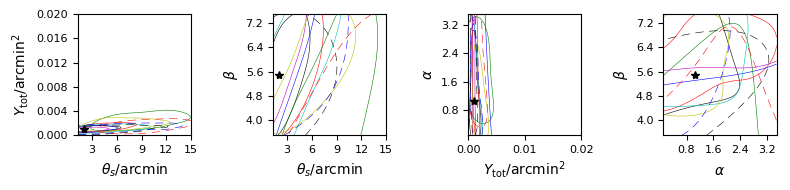} \\
     \includegraphics[width=\textwidth, height = 7.5cm, keepaspectratio]{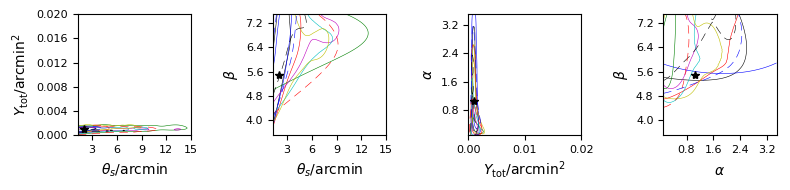} \\
    \end{tabular}
\caption{Two-dimensional marginalised posterior distributions for four combinations of \thetas, \Ytot, \aGNFW\ and \bGNFW, for the 10 UPP low SNR cluster simulations obtained from: AMI data (top row), \textit{Planck} data (middle row), and AMI and \textit{Planck} data combined (bottom row).}
\label{f:ap_obs_lowsnr_vary}
  \end{center}
\end{figure*}

\begin{figure*}
  \begin{center}
    \begin{tabular}{@{}c@{}}
     \includegraphics[width=\textwidth, height = 7.5cm, keepaspectratio]{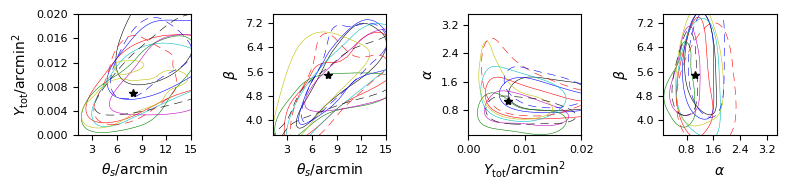} \\
     \includegraphics[width=\textwidth, height = 7.5cm, keepaspectratio]{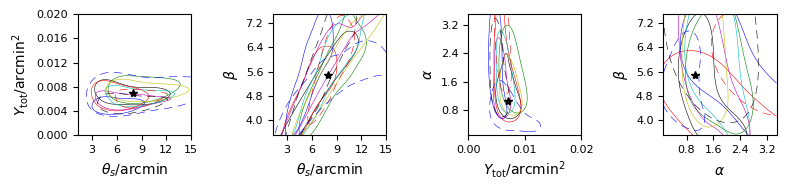} \\
     \includegraphics[width=\textwidth, height = 7.5cm, keepaspectratio]{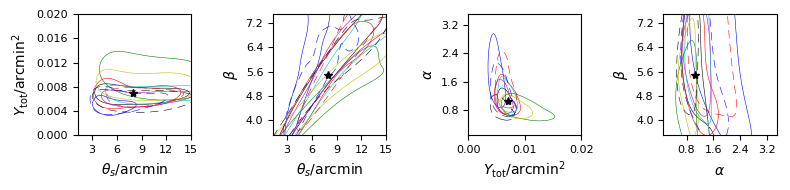} \\
    \end{tabular}
\caption{Two-dimensional marginalised posterior distributions for four combinations of \thetas, \Ytot, \aGNFW\ and \bGNFW, for the 10 UPP high SNR cluster simulations obtained from: AMI data (top row), \textit{Planck} data (middle row), and AMI and \textit{Planck} data combined (bottom row).}
\label{f:ap_obs_highsnr_vary}
  \end{center}
\end{figure*}

Figs.~\ref{f:ap_obs_lowsnr_vary} and \ref{f:ap_obs_highsnr_vary} show two-dimensional posterior constraints for a selection of parameter pairs for the UPP low- and high-SNR cluster simulations respectively.  The two sets of simulations share some common features, as follows.  The constraints on \Ytot\ are mostly driven by the \emph{Planck} data, since the lower-resolution \emph{Planck} data are most suited to measuring the total cluster signal while the interferometric AMI data relies on extrapolations to larger angular scales than are measured; the joint constraints are generally tighter than the \emph{Planck}-only constraints and appear unbiased.  There is a large degeneracy between \thetas\ and \bGNFW\ which is inherent to the GNFW model, since decreasing \bGNFW\ makes the cluster amplitude fall off more slowly with radius which can also be achieved by increasing \thetas.

In the case of the low SNR cluster, the joint analysis successfully constrains \Ytot\ and puts an upper limit on \thetas\ and a lower limit on \bGNFW.  \aGNFW\ is fairly unconstrained since it affects the profile on the scale of \thetas\ = 2\,arcmin which is not well-resolved by either instrument.

In the case of the high SNR cluster, \aGNFW\ can be constrained by AMI alone.  The joint analysis places a lower limit on \thetas\ but \bGNFW\ is unconstrained due to the \thetas/\bGNFW\ degeneracy.

\begin{figure*}
  \begin{center}
    \begin{tabular}{@{}c@{}}
     \includegraphics[width=\textwidth, height = 7.5cm, keepaspectratio]{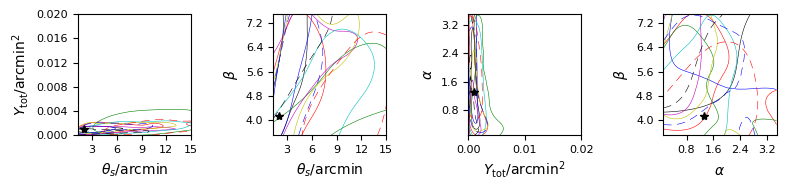} \\
     \includegraphics[width=\textwidth, height = 7.5cm, keepaspectratio]{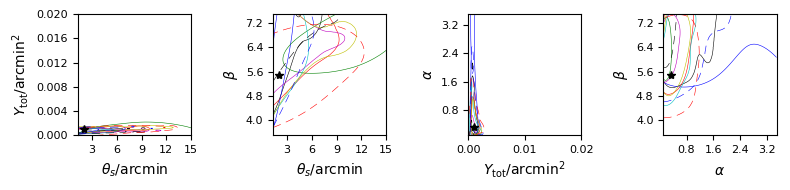} \\
    \end{tabular}
\caption{Two-dimensional marginalised posterior distributions for four combinations of \thetas, \Ytot, \aGNFW\ and \bGNFW, for the 10 PIP low SNR cluster simulations (top row) and REP low SNR cluster simulations (bottom row) obtained from AMI and \textit{Planck} data combined.  These represent our `worst-case' and `best-case' constraints for the set of clusters studied.}
\label{f:pip_rep_vary}
  \end{center}
\end{figure*}

We do not show all of the constraints produced for the PIP and REP simulations as they are qualitatively similar, with the following exceptions.  Since \bGNFW\ is lower for the PIP it effectively makes the cluster much more extended.  The significance of the high SNR AMI detection becomes much lower and with \aGNFW\ and \bGNFW\ allowed to vary all of the parameters are essentially unconstrained, so the joint constraints become driven by the \emph{Planck} data.  \Ytot\ is well-constrained; a lower limit can be put on \thetas\ and an upper limit on \bGNFW, while \aGNFW\ is unconstrained.  In the low SNR case, only \Ytot\ is well-constrained since although the cluster is more extended it is also much less bright so there is not enough signal-to-noise to constrain \aGNFW\ and \bGNFW.  We show the joint constraints for this case in Fig.~\ref{f:pip_rep_vary} as a `worst-case' scenario.

In the high SNR, REP case, \aGNFW\ is very tightly constrained and \bGNFW\ is better-constrained by the AMI data alone.  This is because the lower \aGNFW\ value of the profile causes it to fall off more sharply with radius, putting more signal on AMI scales.  In the low SNR, REP case we obtain tight constraints on \Ytot\ and \aGNFW, a tight upper limit on \thetas\ and a lower limit on \bGNFW.  We show the joint constraints for this case in Fig.~\ref{f:pip_rep_vary} as a `best-case' scenario.  We note that for the REP analysis we are fixing \cGNFW\ to the incorrect, UPP value; we also varied \aGNFW\ and \bGNFW\ while fixing \cGNFW\ to the correct, REP value which had little impact on the parameter constraints.

Overall we conclude that where nothing is known a-priori about the pressure profile of the cluster (other than that it follows a GNFW shape), we can use the combination of AMI and \emph{Planck} data to successfully constrain \Ytot; often constrain \aGNFW; and sometimes put limits on \bGNFW\ and \thetas\ depending on how well-resolved the cluster is (which depends both on \thetas\ and on \aGNFW\ and \bGNFW).

\section{Application of joint analysis to real cluster data}\label{S:real_data}

As a test case, we apply the joint analysis to PSZ2~G063.80+11.42 ($z=0.426$; $M_{500,\mathrm{SZ}} = 6.2 \times 10^{14} M_{\sun}$) from the P15 sample.  The cluster is well-detected by both \emph{Planck} (PwS signal-to-noise-ratio = 6.5) and AMI (Bayesian detection significance = 3.8) and there is a significant offset between the AMI and \emph{Planck} posterior constraints on \Ytot\ and \thetas.  The AMI radio-source environment is relatively clean.  We reobserved the cluster with AMI to benefit from the improved performance of the new correlator, and use the DX11d data release with strong point sources masked (\emph{Planck} Collaboration, priv. comm.) for the \emph{Planck} data.

\begin{figure}
  \begin{center}
     \includegraphics[width=\linewidth]{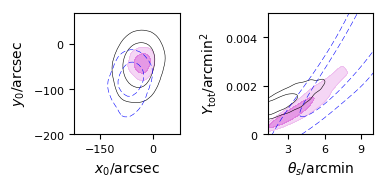} 
     \caption{Posterior distributions for \emph{Planck}-only (black, solid), AMI-only with flat priors (blue, dashed) and AMI-only with P15 priors (magenta, filled contours) analyses of PSZ2~G063.80+11.42, fixing the profile shape parameters at the UPP values.  Note the range of the right-hand plot is truncated to better display the degeneracies.}\label{Fi:CAJ1905_fixedab}
  \end{center}
\end{figure}

We firstly run the AMI and \emph{Planck} analyses separately using the priors given in Table~\ref{t:ap_priors} (assigning delta priors to the GNFW shape parameters at UPP values), to confirm the discrepancy. For comparison we also ran the AMI analysis with the P15 priors; these three sets of posterior chains are shown in Fig.~\ref{Fi:CAJ1905_fixedab}.  We note that the tighter positional constraint in the latter case is due to a degeneracy with a radio source flux near one edge of the cluster; when the wider positional prior is used the source flux is allowed to increase, broadening the cluster decrement and shifting the position of the cluster.  This has little effect on the \thetas/\Ytot\ constraint.  We see that the discrepancy is confirmed with the newer AMI data (and the AMI Bayesian detection significance increases to 6.0); the \emph{Planck} \thetas/\Ytot\ constraint lies significantly above that produced by AMI.

\begin{figure*}
  \begin{center}
     \includegraphics[width=\linewidth]{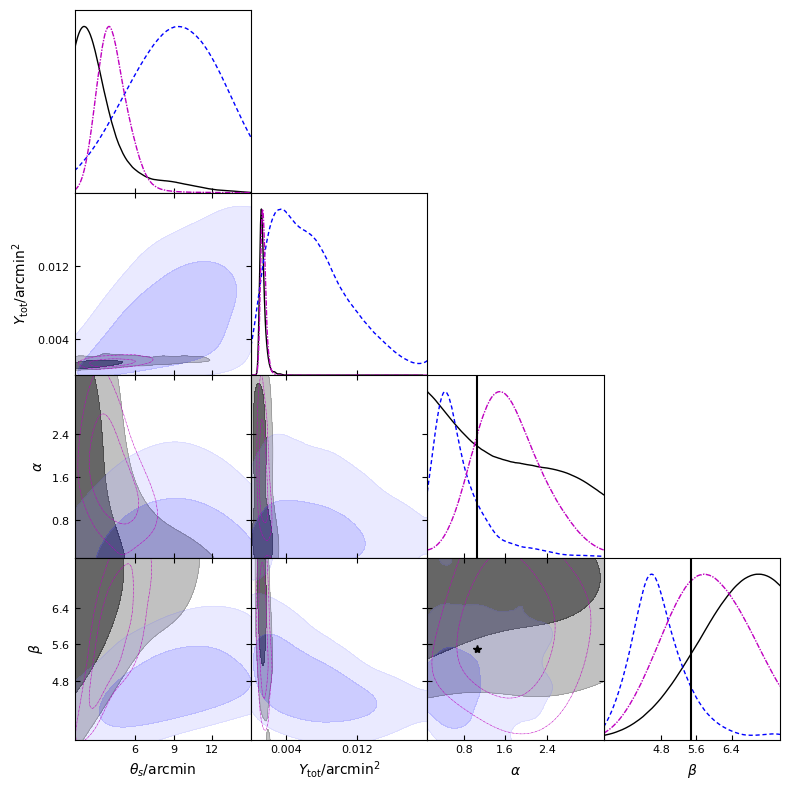} \\\vspace{-\linewidth}
     \hfill\includegraphics[width=0.35\linewidth]{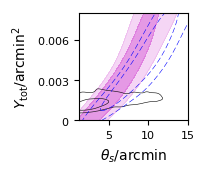}\vspace{0.7\linewidth}
     \caption{Bottom left-hand corner: triangle plot showing posterior distributions for \emph{Planck}-only (black filled contours, solid lines), AMI-only (blue filled contours, dashed lines) and joint analysis (magenta empty contours, dot-dashed lines) of PSZ2~G063.80+11.42, all with the same priors, allowing \aGNFW\ and \bGNFW\ to vary.  The black vertical lines and star show the UPP values of \aGNFW\ and \bGNFW.  In the upper right-hand corner, the blue dashed and filled magenta contours show AMI-only analyses with \aGNFW\ fixed to the UPP value and the joint analysis value (1.66), respectively; the posterior has shifted to better overlap with the \emph{Planck}-only posterior with \aGNFW\ and \bGNFW\ varying, shown with solid black contours.}\label{f:ap_obs_ab_real}
  \end{center}
\end{figure*}

We now analyse the PSZ2~G063.80+11.42 data while allowing \aGNFW\ and \bGNFW\ to vary. Figure~\ref{f:ap_obs_ab_real} shows the resulting posterior distributions for the AMI-only, \emph{Planck}-only and joint analysis methods. Similarly to the simulations, we see that \Ytot\ is well constrained by the joint analysis and the constraint is completely driven by the \emph{Planck} data.  The addition of AMI data improves the constraint on \thetas\ from an upper limit to a true constraint, and both \aGNFW\ and \bGNFW\ are constrained, although not tightly.  The one-dimensional parameter constraints are summarized in Table~\ref{T:CAJ1905_constraints}; \bGNFW\ agrees with the UPP value, while a higher value of \aGNFW\ is favoured but only at the $\approx$\,$1\sigma$ level.  We perform one further analysis on the AMI data, fixing \aGNFW\ to the fitted value from the joint analysis, and leaving \bGNFW\ fixed to the UPP value.  The posterior constraints on \thetas\ and \Ytot\ in this case are shown in comparison with the AMI-only UPP analysis and the \emph{Planck} analysis varying \aGNFW\ and \bGNFW; it can be seen that the AMI posterior shifts in the correct direction to overlap with \emph{Planck}, confirming that this value of \aGNFW\ gives better agreement between the two instruments (Fig.~\ref{f:ap_obs_ab_real}, upper right-hand corner). No X-ray observations of this cluster are available to compare X-ray-derived constraints on the profile parameters; for our future AMI-\emph{Planck} sample (see Section~\ref{S:future_work}) comparison with X-ray-derived parameters will be informative and complementary across a different range of angular scales.

\begin{table}
\begin{center}
\caption{Summary of parameter constraints for PSZ2~G063.80+11.42, varying \aGNFW\ and \bGNFW.  The errors given are the 68\% limits from the one-dimensional marginalised parameter constraints.}
\label{T:CAJ1905_constraints}
\begin{tabular}{lccc}
\hline
 & AMI-only & \emph{Planck}-only & AMI-\emph{Planck} \\ \hline
\xoff / arcsec & $-59^{+24}_{-18}$ & $-38^{+31}_{-26}$ & $-55\pm 14$ \\[3pt]
\yoff / arcsec & $-70^{+26}_{-14}$ & $-45\pm 31$ & $-65\pm 15$ \\[3pt]
\Ytot / ($10^3$ arcmin$^{2}$) & $6.9^{+2.3}_{-5.7}$ & $1.32^{+0.17}_{-0.38}$ & $1.40^{+0.23}_{-0.34}$ \\[3pt]
\thetas / arcmin & $8.9^{+3.9}_{-3.0}$ & $< 3.81$ & $4.27^{+0.95}_{-1.4}$ \\[3pt]
\aGNFW & $0.77^{+0.10}_{-0.63}$ & $< 2.16$ & $1.66^{+0.55}_{-0.74}$ \\[3pt]
\bGNFW &$4.79^{+0.42}_{-0.74}$ & $> 5.94$ & $5.78^{+1.0}_{-0.87}$ \\[3pt]
\hline
\end{tabular}
\end{center}
\end{table}


\section{Future work}\label{S:future_work}

Along with the sample of 99 clusters from P15, clusters which were previously excluded from the AMI sample due to difficult radio source environments are currently being reobserved with the new correlator; the superior dynamic range of the new instrument allows us to cope better with these environments and successfully extract cluster parameters.  We will analyse all AMI detections with the joint pipeline, giving us a large cluster sample to probe deviations from the UPP and consider the impact this may have on the \emph{Planck} cluster number counts.  

We note that \citet{2019MNRAS.483.3529J} found that realistic radio source environments could bias the recovery of cluster parameters from AMI data.  This issue and its effect on the recovery of the pressure profile parameters will be investigated in conjunction with the analysis of the larger AMI-\emph{Planck} sample.

With the larger sample we will be able to compare our SZ-derived profile parameters to X-ray-derived parameters.  This will allow us to test for any systematic differences and combine information across a broader range of angular scales.

Our pipeline is also simply extensible to the use of any physical model which uses a GNFW profile for the gas pressure, e.g.\ the models proposed in \citet{2012MNRAS.423.1534O}, \citet{2018arXiv180501968J} and \citet{2018arXiv180903325J}.  We also plan to implement a non-parametric model such as that proposed in \citet{2018MNRAS.481.3853O}.


\section{Conclusions}
\label{S:conclusions}

\begin{enumerate}
\item{We have developed a joint likelihood function for SZ data obtained from the \textit{Planck} satellite and the Arcminute Microkelvin Imager (AMI) radio interferometer system, in order to compare inferences obtained using it with those from the individual likelihoods. The method could apply to any combination of \emph{Planck} and interferometric data from one or more telescopes.}

\item{We generated simulations of clusters using an observational model similar to the one used in \citet{2015A&A...580A..95P}, using gas pressure profile shape parameter values taken from either the UPP (\citealt{2010A&A...517A..92A}, `universal' pressure profile) or two other realistic variations. We considered both a smaller angular size, fainter cluster and a larger angular size, brighter cluster. From looking at the resulting posterior distributions we found the following:
\begin{itemize}
\item When \textit{simulating} and \textit{analysing} the clusters using the model with UPP parameters the joint analysis greatly reduced the degeneracy in \Ytot-\thetas\ shown in the individual AMI and \emph{Planck} analyses, due to the different degeneracy directions for the individual datasets. The improvement on the parameter constraints for the joint analysis is particularly prominent in the small angular size cases.  Thus when the profile shape of a cluster is known a-priori, the combination of the two datasets provides accurate, precise constraints on the cluster parameters with no need for external information to reduce the \Ytot-\thetas\ degeneracy.
\item When simulated clusters created using non-UPP profiles are analysed with the profile shape assumed to be UPP, the \Ytot-\thetas\ constraints are biased away from the true value.  This occurs in the individual AMI and \emph{Planck} datasets and is particularly problematic in the joint analysis, where a tight, significantly biased constraint is produced.
\item When allowing the shape parameters to vary in the Bayesian analysis, we generally found that for all the clusters \Ytot\ was well-constrained and unbiased; \emph{Planck} drove the constraint and the joint analyses improved the constraint slightly.
\item Furthermore dependent on how well resolved the clusters are, the shape parameter \aGNFW\ can often be constrained.  Due to the strong \bGNFW-\thetas\ degeneracy these parameters are more difficult to constrain and it is usually only possible to place limits on them.
\end{itemize}
}
\item{Finally, we applied the joint analysis to real data for the cluster PSZ2~G063.80+11.42 which is part of the sample of 99 clusters considered in P15.  We confirmed the discrepancy in \Ytot\ and \thetas\ estimates when using updated AMI data and resolved it by allowing \aGNFW\ and \bGNFW\ to vary.  Using the joint analysis we could constrain \Ytot\ and \thetas\ well and place loose constraints on \aGNFW\ and \bGNFW, finding that a slightly higher value of \aGNFW\ than the UPP value was preferred, while the constraint on \bGNFW\ was consistent with the UPP value.}
\item{We plan to apply our method to all the 99 clusters of the P15 sample, plus clusters currently being reobserved by AMI, to investigate deviations from the UPP and possible impact on the \emph{Planck} cluster number counts.  We also plan to extend our method to incorporate different physical and non-parametric cluster models.}
\end{enumerate}


\section*{Acknowledgements}
We thank the staff of the Mullard Radio Astronomy Observatory for their invaluable assistance in the commissioning and operation of AMI, which is supported by Cambridge and Oxford Universities.  This work is based on observations obtained with Planck (http://www.esa.int/Planck), an ESA science mission with instruments and contributions directly funded by ESA Member States, NASA, and Canada.  This work was performed using the Darwin Supercomputer of the University of Cambridge High Performance Computing (HPC) Service (\url{http://www.hpc.cam.ac.uk/}), provided by Dell Inc. using Strategic Research Infrastructure Funding from the Higher Education Funding Council for England and funding from the Science and Technology Facilities Council. The authors would like to thank Stuart Rankin from HPC and Greg Willatt and David Titterington from Cavendish Astrophysics for computing assistance. KJ and PJE acknowledge Science and Technology Facilities Council studentships. YCP acknowledges support from a Trinity College JRF and Rutherford Discovery Fellowship.


\setlength{\bibsep}{0pt}            
\renewcommand{\bibname}{References} 


\bsp	
\label{lastpage}
\end{document}